\documentclass[12pt,aps,showpacs,notitlepage]{revtex4-1}
\usepackage{amssymb,amsmath}
\usepackage{bm}
\usepackage[pdftex]{graphicx}
\usepackage{graphicx}
\usepackage{epstopdf}

\usepackage{subfig}
\usepackage{color} 

\newcommand{\beq}{\begin{equation}}
\newcommand{\beqn}{\begin{eqnarray}}
\newcommand{\eeq}{\end{equation}}
\newcommand{\eeqn}{\end{eqnarray}}
\begin{document}

\title{Multifield Inflation after {\it Planck}: \\ The Case for Nonminimal Couplings}
\author{David I. Kaiser and Evangelos I. Sfakianakis}
\email{Email addresses: dikaiser@mit.edu; esfaki@mit.edu}
\affiliation{Center for Theoretical Physics and Department of Physics, \\
Massachusetts Institute of Technology, Cambridge, Massachusetts 02139 USA}
\date{\today}
\begin{abstract} Multifield models of inflation with nonminimal couplings are in excellent agreement with the recent results from {\it Planck}. Across a broad range of couplings and initial conditions, such models evolve along an effectively single-field attractor solution and predict values of the primordial spectral index and its running, the tensor-to-scalar ratio, and non-Gaussianities squarely in the observationally most-favored region. Such models also can amplify isocurvature perturbations, which could account for the low power recently observed in the CMB power spectrum at low multipoles. Future measurements of primordial isocurvature perturbations could distinguish between the currently viable possibilities.
\end{abstract}
\pacs{04.62+v; 98.80.Cq. Published in {\it Phys. Rev. Lett.} {\bf 112} (2014): 011302.}
\maketitle


Early-universe inflation remains the leading framework for understanding a variety of features of our observable universe \cite{GuthKaiser,BTW}. Most impressive has been the prediction of primordial quantum fluctuations that could seed large-scale structure. Recent measurements of the spectral tilt of primordial (scalar) perturbations, $n_s$, find a decisive departure from a scale-invariant spectrum \cite{WMAP9,PlanckInflation}. The {\it Planck} collaboration's value, $n_s = 0.9603 \pm 0.0073$, differs from $n_s = 1$ by more than $5\sigma$. At the same time, observations with {\it Planck} constrain the ratio of tensor-to-scalar perturbations to $r < 0.11 \> (95\% \> {\rm CL})$, and are consistent with the absence of primordial non-Gaussianities, $f_{\rm NL} \sim 0$ \cite{PlanckInflation,PlanckNonGauss}. 

The {\it Planck} team also observes less power in the angular power spectrum of temperature anisotropies in the cosmic microwave background radiation (CMB) at low multipoles, $\ell \sim 20 - 40$, compared to best-fit $\Lambda$CDM cosmology: a $2.5 - 3\sigma$ departure on large angular scales, $\theta > 5^\circ$ \cite{PlanckCMB}. Many physical processes might ultimately account for the deviation, but a primordial source seems likely given the long length-scales affected. One plausible possibility is that the discrepancy arises from the amplification of isocurvature modes during inflation \cite{PlanckInflation}. 

In this brief paper we demonstrate that simple, well-motivated multifield models with nonminimal couplings match the latest observations particularly well, with no fine-tuning. This class of models (i) generically includes potentials that are concave rather than convex at large field values, (ii) generically predicts values of $r$ and $n_s$ in the most-favored region of the recent observations. (iii) generically predicts $f_{\rm NL} \sim 0$ except for exponentially fine-tuned initial field values, (iv) generically predicts ample entropy production at the end of inflation, with an effective equation of state $w_{\rm eff} \sim [0, 1/3]$, and (v) generically includes isocurvature perturbations as well as adiabatic perturbations, which might account for the low power in the CMB power spectrum at low multipoles. 

We consider this class of models to be well-motivated for several reasons. Realistic models of particle physics include multiple scalar fields at high energies. In any such model, nonminimal couplings are {\it required} for self-consistency, since they arise as renormalization counterterms when quantizing scalar fields in curved spacetime \cite{BirrellDavies}. Moreover, the nonminimal coupling constants generically rise with energy under renormalization-group flow with no UV fixed-point \cite{Buchbinder}, and hence one expects $\vert \xi \vert \gg 1$ at inflationary energy scales. In such models inflation occurs for field values and energy densities well below the Planck scale (see \cite{KMS,GKS,HiggsInflation} and references therein). Higgs inflation \cite{HiggsInflation} is an elegant example: in renormalizable gauges (appropriate for high energies) the Goldstone modes remain in the spectrum, yielding a multifield model \cite{GKS,Goldstones1,Goldstones2}.

We demonstrate here for the first time that models of this broad class exhibit an attractor behavior: over a wide range of couplings and fields' initial conditions, the fields evolve along an effectively single-field trajectory for most of inflation. Although attractor behavior is common for single-field models of inflation \cite{SalopekBond1990}, the dynamics of multifield models generally show strong sensitivity to couplings and initial conditions (see, e.g., \cite{Easther} and references therein). Not so for the class of multifield models examined here, thanks to the shape of the effective potential in the Einstein frame. The multifield attractor behavior demonstrated here means that for most regions of phase space and parameter space, this general class of models yields values of $n_s$, $r$, the running of the spectral index $\alpha = d n_s / d \ln k$, and $f_{\rm NL}$ in excellent agreement with recent observations. The well-known empirical success of single-field models with nonminimal couplings \cite{singlefield,HiggsInflation} is thus preserved for more realistic models involving multiple fields. Whereas the attractor behavior creates a large observational degeneracy in the $r$ vs. $n_s$ plane, the isocurvature spectra from these models depend sensitively upon couplings and initial conditions. Future measurements of primordial isocurvature spectra could therefore distinguish among models in this class.


In the Jordan frame, the fields' nonminimal couplings remain explicit in the action,
\beq
S_{\rm J} = \int d^4 x \sqrt{-\tilde{g}} \left[ f (\phi^I ) \tilde{R} - \frac{1}{2} \delta_{IJ} \tilde{g}^{\mu\nu} \partial_\mu \phi^I \partial_\nu \phi^J - \tilde{V} (\phi^I ) \right] ,
\label{SJ}
\eeq
where quantities in the Jordan frame are marked by a tilde. Performing the usual conformal transformation, $\tilde{g}_{\mu\nu} (x) \rightarrow g_{\mu\nu} (x) = 2 M_{\rm pl}^{-2} f (\phi^I (x)) \> \tilde{g}_{\mu\nu} (x)$, where $M_{\rm pl} \equiv (8 \pi G)^{-1/2} = 2.43 \times 10^{18}$ GeV is the reduced Planck mass, we may write the action in the Einstein frame as \cite{KMS}
\beq
S_{\rm E} = \int d^4 x \sqrt{-g} \left[ \frac{M_{\rm pl}^2}{2} R - \frac{1}{2} {\cal G}_{IJ} g^{\mu\nu} \partial_\mu \phi^I \partial_\nu \phi^J - V (\phi^I ) \right] .
\label{SE}
\eeq
The potential in the Einstein frame, $V (\phi^I )$, is stretched by the conformal factor compared to the Jordan-frame potential:
\beq
V (\phi^I ) = \frac{M_{\rm pl}^4}{4 f^2 (\phi^I ) } \tilde{V} (\phi^I ) .
\label{VE}
\eeq
The nonminimal couplings induce a curved field-space manifold in the Einstein frame with metric ${\cal G}_{IJ} (\phi^K) = (M_{\rm pl}^2 / (2f )) [ \delta_{IJ} + 3 f_{, I} f_{, J} / f  ]$, where $f_{, I} = \partial f / \partial \phi^I $ \cite{KMS}. We adopt the form for $f (\phi^I)$ required for renormalization \cite{BirrellDavies},
\beq
f (\phi^I ) = \frac{1}{2} \left[ M_{\rm pl}^2 + \sum_I \xi_I (\phi^I )^2 \right] .
\label{f}
\eeq
Here we consider two-field models, $I, J = \phi, \chi$.

As emphasized in \cite{HiggsInflation,GKS,KMS}, the conformal stretching of the Einstein-frame potential, Eq. (\ref{VE}), generically leads to concave potentials at large field values, even for Jordan-frame potentials that are convex. In particular, for a Jordan-frame potential of the simple form $\tilde{V} (\phi^I ) = \frac{\lambda_\phi}{4} \phi^4 + \frac{g}{2} \phi^2 \chi^2 + \frac{\lambda_\chi}{4} \chi^4$, Eqs. (\ref{VE}) and (\ref{f}) yield a potential in the Einstein frame that is nearly flat for large field values, $V (\phi^I) \rightarrow \lambda_J M_{\rm pl}^4  / (4 \xi_J^2 )$ (no sum on $J$), as the $J$th component of $\phi^I$ becomes arbitrarily large. This basic feature leads to ``extra"-slow-roll evolution of the fields during inflation. If the couplings $\lambda_J$ and $\xi_J$ are not equal to each other, $V (\phi^I)$ develops ridges separated by valleys  \cite{KMS}. Inflation occurs in the valleys as well as along the ridges, since both are regions of false vacuum with $V \neq 0$. See Fig. 1.

\begin{figure}
\centering
\includegraphics[width=3in]{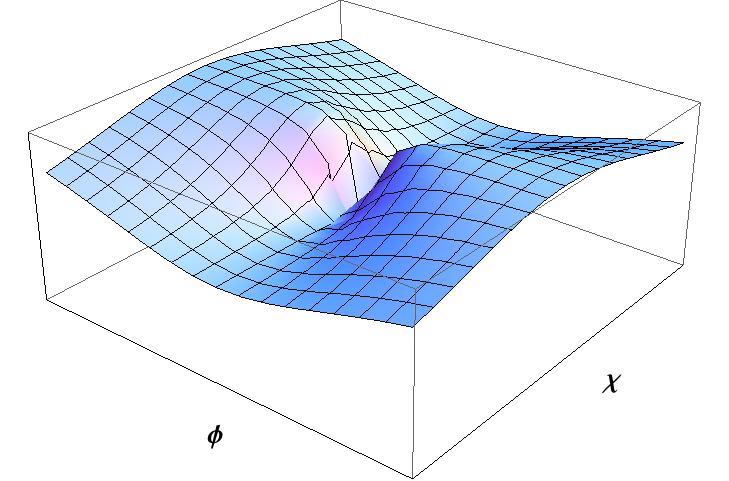} \includegraphics[width=3in]{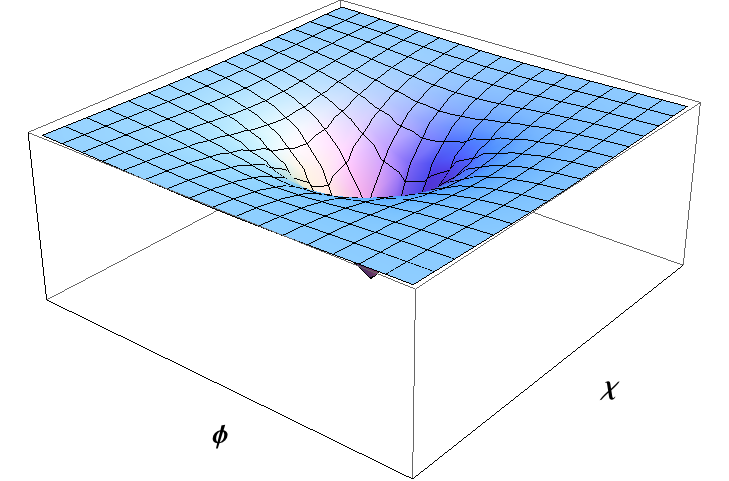}
\caption{\small \baselineskip 14pt Potential in the Einstein frame, $V (\phi^I )$. {\it Left}:  $\lambda_\chi = 0.75 \> \lambda_\phi$, $g = \lambda_\phi$, $\xi_\chi = 1.2 \> \xi_\phi$. {\it Right}: $\lambda_\chi = g = \lambda_\phi$, $\xi_\phi = \xi_\chi$. In both cases, $\xi_I \gg 1$ and $0 < \lambda_I , g < 1$.}
\label{VEVHiggs}
\end{figure}

Constraints on $r$ constrain the energy scale of inflation, $H (t_* ) / M_{\rm pl} < 3.7 \times 10^{-5}$ \cite{PlanckInflation}. For Higgs inflation, with $\lambda_I = g = \lambda_\phi$ and $\xi_I = \xi_\phi$, the Hubble parameter during slow roll is given by $H / M_{\rm pl} \simeq \sqrt{ \lambda_\phi / (12 \xi_\phi^2 )}$. Measurements of the Higgs mass near the electroweak symmetry-breaking scale require $\lambda_\phi \simeq 0.13$. Under renormalization-group flow, $\lambda_\phi$ will fall to the range $0 < \lambda_\phi < 0.01$ at the inflationary energy scale; $\lambda_\phi = 0.01$ requires $\xi_\phi \geq 780$ to satisfy the constraint on $H (t_* ) / M_{\rm pl}$, which in turn requires $\xi_\phi \sim {\cal O} (10^1 - 10^2)$ at low energies \cite{HiggsRunning}. For our general class of models, we therefore consider couplings at the inflationary energy scale of order $\lambda_I , g \sim {\cal O} (10^{-2})$ and $\xi_I \sim {\cal O} (10^3)$ \cite{unitarityfootnote}.

Expanding the scalar fields to first order, $\phi^I (x^\mu) = \varphi^I (t) + \delta \phi^I (x^\mu)$, we find \cite{KMS,GKS}
\beq
\dot{\sigma}^2 =  {\cal G}_{IJ} \dot{\varphi}^I \dot{\varphi}^J  = \left( \frac{M_{\rm pl}^2}{2f} \right) \left[ \dot{\phi}^2 + \dot{\chi}^2 + \frac{3 \dot{f}^2}{f} \right].
\label{sigma}
\eeq
We also expand the spacetime metric to first order around a spatially flat Friedmann-Robertson-Walker metric. Then the background dynamics are given by \cite{KMS}
\beq
\begin{split}
H^2 = \frac{1}{ 3 M_{\rm pl}^2} & \left[ \frac{1}{2} \dot{\sigma}^2 + V \right] , \>\>\> \dot{H} = - \frac{1}{2 M_{\rm pl}^2 } \dot{\sigma}^2 , \\
{\cal D}_t \dot{\varphi}^I &+ 3 H \dot{\varphi}^I + {\cal G}^{IK} V_{, K} = 0 ,
\end{split}
\label{Friedmann}
\eeq
where ${\cal D}_t$ is the (covariant) directional derivative, ${\cal D}_t A^I \equiv \dot{\varphi}^J {\cal D}_J A^I = \dot{A}^I + \Gamma^I_{\> JK} A^J \dot{\varphi}^K$ \cite{PTGeometric,KMS}. The gauge-invariant Mukhanov-Sasaki variables for the linearized perturbations, $Q^I$, obey an equation of motion with a mass-squared matrix given by \cite{PTGeometric,KMS}
\beq
{\cal M}^I_{\>\> J} \equiv {\cal G}^{IJ} \left( {\cal D}_J {\cal D}_K V \right) - {\cal R}^I_{\>\> LMJ} \dot{\varphi}^L \dot{\varphi}^M ,
\label{MIJ}
\eeq
where ${\cal R}^I_{\>\> LMJ}$ is the Riemann tensor for the field-space manifold. 

To analyze inflationary dynamics, we use a multifield formalism (see \cite{BTW,WandsReview} for reviews) made covariant with respect to the nontrivial field-space curvature (see \cite{KMS,PTGeometric} and references therein). We define adiabatic and isocurvature directions in the curved field space via the unit vectors $\hat{\sigma}^I \equiv \dot{\varphi}^I /\dot{\sigma}$ and $\hat{s}^I \equiv \omega^I / \omega$, where the turn-rate vector is given by $\omega^I \equiv {\cal D}_t \hat{\sigma}^I$, and $\omega = \vert \omega^I \vert$. We also define slow-roll parameters \cite{KMS,PTGeometric}:
\beq
\epsilon \equiv - \frac{\dot{H}}{H^2} , \>\>
\eta_{\sigma\sigma} \equiv M_{\rm pl}^2 \frac{ \hat{\sigma}_I \hat{\sigma}^J {\cal M}^I_{\> \> J} }{ V} , \> \> 
\eta_{ss} \equiv M_{\rm pl}^2 \frac{ \hat{s}_I \hat{s}^J {\cal M}^I_{\>\> J} }{V} .
\label{slowroll}
\eeq
Using Eq. (\ref{Friedmann}), we have the exact relation, $\epsilon = 3 \dot{\sigma}^2 /( \dot{\sigma}^2 + 2 V )$. The adiabatic and isocurvature perturbations may be parameterized as ${\cal R}_c = (H / \dot{\sigma} ) \hat{\sigma}_I Q^I$ and ${\cal S} = ( H / \dot{\sigma} ) \hat{s}_I Q^I$, where ${\cal R}_c$ is the gauge-invariant curvature perturbation. Perturbations of pivot-scale $k_* = 0.002 \> {\rm Mpc}^{-1}$ first crossed outside the Hubble radius during inflation at time $t_*$. In the long-wavelength limit, the evolution of ${\cal R}_c$ and ${\cal S}$ for $t > t_*$ is given by the transfer functions \cite{KMS,PTGeometric}
\beq
\begin{split}
T_{\cal RS} (t_*, t) &= \int_{t_*}^t dt' \> 2 \omega (t') T_{\cal SS} (t_*, t') , \\
T_{\cal SS} (t_* , t) &= {\rm exp} \left[ \int_{t_*}^t dt' \> \beta (t') H (t') \right],
\end{split}
\label{TRSTSS}
\eeq
with $\beta (t) = - 2 \epsilon - \eta_{ss} + \eta_{\sigma\sigma} - \frac{4}{3} \frac{\omega^2}{H^2}$. Given the form of $T_{\cal RS}$, the perturbations ${\cal R}_c$ and ${\cal S}$ decouple if $\omega^I = 0$. 

The dimensionless power spectrum for the adiabatic perturbations is defined as ${\cal P}_{\cal R} (k) = (2\pi)^{-2} k^3 \vert {\cal R}_c \vert^2$ and the spectral index is defined as $n_s - 1 \equiv \partial \ln {\cal P}_{\cal R} / \partial \ln k$. Around $t_*$ the spectral index is given by \cite{KMS,BTW,WandsReview,PTGeometric}
\beq
n_s (t_*) = 1 - 6 \epsilon (t_*) + 2 \eta_{\sigma\sigma} (t_*) .
\label{nsstar}
\eeq
At late times and in the long-wavelength limit, the power spectrum becomes ${\cal P}_{\cal R}  = {\cal P}_{\cal R} (k_*) \left[ 1 + T_{\cal RS}^2  \right] $, and hence the spectral index may be affected by the transfer of power from isocurvature to adiabatic modes: $n_s (t) = n_s (t_*) + H_*^{-1} (\partial T_{\cal RS} / \partial t_* ) \sin (2 \Delta )$, with $\cos \Delta \equiv T_{\cal RS} ( 1 + T_{\cal RS}^2 )^{-1/2}$. 

The mass of the isocurvature perturbations is $\mu_s^2 = 3 H^2 (\eta_{ss} + {\omega^2 / H^2} )$ \cite{KMS}. For $\mu_s < 3 H / 2$ we have ${\cal P}_{\cal S} (k_*) \simeq {\cal P}_{\cal R} (k_*)$ and hence ${\cal P}_S \simeq {\cal P}_{\cal R} (t_*) T_{\cal SS}^2$ at late times. In the Einstein frame the anisotropic pressure $\Pi^i_{\> \> j} \propto T^i_{\>\> j}$ for $i \neq j$ vanishes to linear order, so the tensor perturbations $h_{ij}$ evolve just as in single-field models with ${\cal P}_T \simeq 128  \>  [ H (t_*) / M_{\rm pl} ]^2 (k / k_*)^{- 2 \epsilon}$, and therefore $r \equiv {\cal P}_T / {\cal P}_{\cal R} = 16 \epsilon / [1 + T_{\cal RS}^2 ]$ \cite{BTW,WandsReview,PTGeometric}.


To study the single-field attractor behavior, we first consider the case in which the system inflates in a valley along the $\chi = 0$ direction, perhaps after first rolling off a ridge. In the slow-roll limit and with $\chi \sim \dot{\chi} \sim 0$, Eq. (\ref{Friedmann}) reduces to \cite{GKS}
\beq
\dot{\phi}_{\rm SR} \simeq - \frac{ \sqrt{\lambda_\phi} \> M_{\rm pl}^3 }{ 3 \sqrt{3} \> \xi_\phi^2 \phi} .
\label{phidotSR}
\eeq
Using $H / M_{\rm pl} \simeq \sqrt{\lambda_\phi / (12 \xi_\phi^2)}$ we may integrate Eq. (\ref{phidotSR}),
\beq
\frac{\xi_\phi \phi_*^2}{M_{\rm pl}^2} \simeq \frac{4}{3} N_* ,
\label{Nstar}
\eeq
where $N_*$ is the number of efolds before the end of inflation, and we have used $\phi (t_*) \gg \phi (t_{\rm end})$. (We arrive at comparable expressions if the system falls into a valley along some angle in field space, $\theta \equiv {\rm arctan} \> ( \phi / \chi )$.) Eq. (\ref{sigma}) becomes $\dot{\sigma}^2\vert_{\chi = 0} \simeq 6 M_{\rm pl}^2 \dot{\phi}^2  / \phi^2 $ upon using $\xi_\phi \gg 1$. Using $V \simeq \lambda_\phi M_{\rm pl}^4  / (4 \xi_\phi^2)$ and Eqs. (\ref{phidotSR}), (\ref{Nstar}) in Eq. (\ref{slowroll}) we find
\beq
\epsilon \simeq \frac{3}{4 N_*^2 } .
\label{epsilonapprox}
\eeq
To estimate $\eta_{\sigma\sigma}$ we use $\eta_{\sigma\sigma} = \epsilon - \ddot{\sigma} / (H \dot{\sigma} ) + {\cal O} (\epsilon^2)$ \cite{BTW}, and find 
\beq
\eta_{\sigma\sigma} \simeq - \frac{1}{ N_* } \left( 1 - \frac{3}{ 4 N_* } \right).
\label{etasigmasigmaapprox}
\eeq

All dependence on $\lambda_I$ and $\xi_I$ has dropped out of these expressions for $\epsilon$ and $\eta_{\sigma\sigma}$ in Eqs. (\ref{epsilonapprox}) and (\ref{etasigmasigmaapprox}). For a broad range of initial field values and velocities --- and {\it independent} of the couplings --- this entire class of models should quickly relax into an attractor solution in which the fields evolve along an effectively single-field trajectory with vanishing turn-rate, $\omega^I \sim 0$. Within this attractor solution we find analytically $\epsilon_* = 2.08 \times 10^{-4}$ and $\eta_{\sigma\sigma *} = - 0.0165$ for $N_* = 60$; and $\epsilon_* = 3.00 \times 10^{-4}$ and $\eta_{\sigma\sigma *}  = - 0.0197$ for $N_* = 50$. To test this attractor behavior, we performed numerical simulations with a sampling of couplings and initial conditions. We fixed $\lambda_\phi = 0.01$ and $\xi_\phi = 10^3$ and looped over $\lambda_\chi = \{ 0.5, 0.75 , 1 \} \> \lambda_\phi$, $g = \{ 0.5, 0.75 , 1\}  \> \lambda_\phi$, and $\xi_\chi = \{ 0.8 , 1 , 1.2 \}  \> \xi_\phi $. These parameters gave a variety of potentials with combinations of ridges and valleys along different directions in field space. We set the initial amplitude of the fields to be $\sqrt{ \phi_0^2  + \chi_0^2 } = 10 \times {\rm max} \> [\xi_\phi^{-1/2} , \xi_\chi^{-1/2} ]$ (in units of $M_{\rm pl}$), which generically produced 70 or more efolds of inflation. We varied the initial angle in field space, $\theta_0 = {\rm arctan} \> ( \phi_0 / \chi_0 )$, among the values $\theta_0 = \{ 0, \pi / 6 , \pi / 3 , \pi / 2 \}$, and allowed for a relatively wide range of initial fields velocities: $\dot{\phi}_0 , \dot{\chi}_0 = \{ - 10 \> \vert \dot{\phi}_{\rm SR} \vert , 0 , +10 \> \vert \dot{\phi}_{\rm SR}  \vert \}$, where $\dot{\phi}_{\rm SR}$ is given by Eq. (\ref{phidotSR}). 

\begin{figure}
\centering
\includegraphics[width=3in]{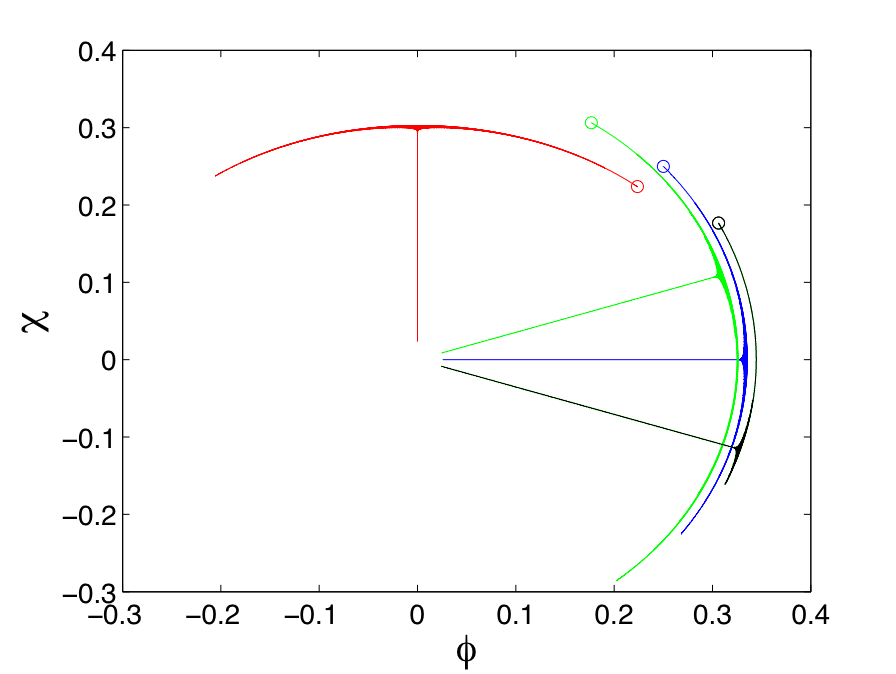} \includegraphics[width=3in]{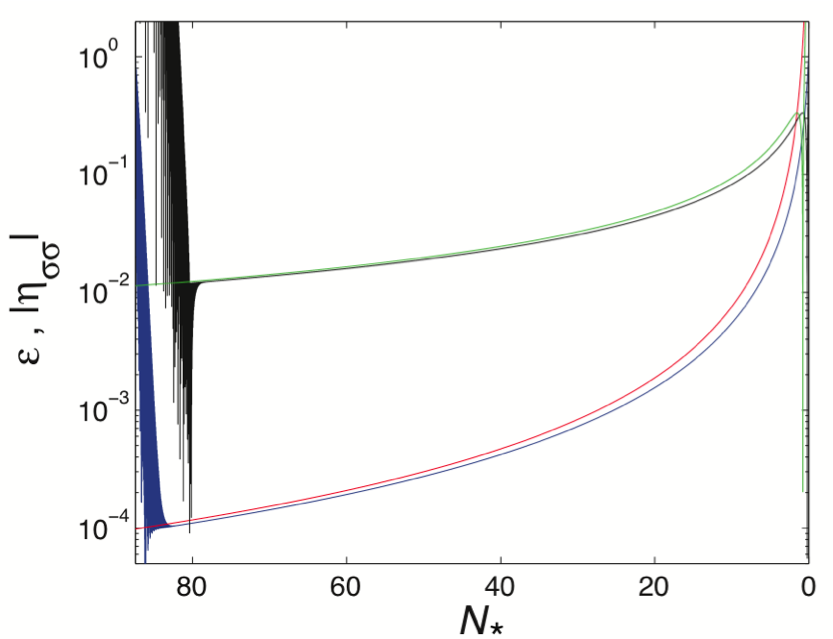} 
\caption{\small \baselineskip 14pt {\it Left}: Field trajectories for different couplings and initial conditions (here for $\dot{\phi}_0 , \dot{\chi}_0 = 0$). Open circles indicate fields' initial values. The parameters $\{\lambda_\chi, g,\xi_\chi, \theta_0 \}$ are given by: $\{0.75 \> \lambda_\phi,  \lambda_\phi, 1.2\> \xi_\phi, \pi/4 \} $ (red),
$\{\lambda_\phi, \lambda_\phi, 0.8 \> \xi_\phi, \pi/4 \} $ (blue), $ \{\lambda_\phi, 0.75 \> \lambda_\phi, 0.8\xi_\phi, \pi/6 \} $ (green), $ \{\lambda_\phi,  0.75 \> \lambda_\phi, 0.8\xi_\phi, \pi/3 \} $ (black).  {\it Right}: Numerical vs. analytic evaluation of the slow-roll parameters, $\epsilon$ (numerical = blue, analytic = red) and $\eta_{\sigma\sigma}$ (numerical = black, analytic = green), for $\lambda_\phi = 0.01$, $\lambda_\chi = 0.75 \> \lambda_\phi$, $g = \lambda_\phi$, $\xi_\phi = 10^3$, and $\xi_\chi = 1.2 \> \xi_\phi$, with $\theta_0 = \pi / 4$ and $\dot{\phi}_0 = \dot{\chi}_0 = +10 \> \vert \dot{\phi}_{\rm SR} \vert$.    }
\label{trajectories}
\end{figure}

Typical trajectories are shown in Fig. \ref{trajectories}a. In each case, the fields quickly rolled into a valley and, after a brief, transient period of oscillation, evolved along a straight trajectory in field-space for the remainder of inflation with $\omega^I = 0$. Across this entire range of couplings and initial conditions, the analytic expressions for $\epsilon$ and $\eta_{\sigma\sigma}$ in Eqs. (\ref{epsilonapprox})-(\ref{etasigmasigmaapprox}) provide close agreement with the exact numerical simulations. See Fig. 2b.

We confirmed numerically that for much larger initial field velocities, up to $\dot{\phi}_0, \dot{\chi}_0 \sim 10^6 \vert \dot{\phi}_{\rm SR} \vert$, such that the initial kinetic energy is larger than the difference between ridge-height and valley in the potential, the system exhibits a very brief, transient period of rapid angular motion (akin to \cite{GKS}). The fields' kinetic energy rapidly redshifts away so that the fields land in a valley of the potential within a few efolds, after which slow-roll inflation continues along a single-field attractor trajectory just like the ones shown in Fig. \ref{trajectories}a. Moreover, the attractor behavior is unchanged if one considers bare masses $m_\phi, m_\chi \ll M_{\rm pl}$ or a negative coupling $g < 0$, so long as one imposes the fairly minimal constraint that $V \geq 0$ and hence $g > - \sqrt{\lambda_\phi \lambda_\chi}$. (Each of these features could affect preheating dynamics but not the attractor behavior during inflation.) Lastly, we performed numerical simulations for the case of three fields rather than two, and again found that the dynamics quickly relax to the single-field attractor since the effective potential contains ridges and valleys, so the fields generically wind up within a valley.


As we have confirmed numerically, trajectories in the single-field attractor solution generically have $\omega^I \sim 0$ between $t_*$ and $t_{\rm end}$ (which we define as $\epsilon (t_{\rm end}) = 1$, or $\ddot{a} (t_{\rm end}) = 0$); hence $T_{\cal RS} \sim 0$ for these trajectories. The spectral index $n_s (t)$ therefore reduces to $n_s (t_*)$ of Eq. (\ref{nsstar}), and $r$ reduces to $r = 16 \epsilon \> [1 + {\cal O} (T_{\cal RS}^2 ) ] \simeq 16 \epsilon$. Using Eqs. (\ref{epsilonapprox}) and (\ref{etasigmasigmaapprox}), we then find
\beq
n_s  \simeq 1 - \frac{2}{N_*} - \frac{3}{N_*^2} , \> \> r \simeq \frac{12}{N_*^2} ,
\label{nsr}
\eeq
and hence $n_s = 0.966$ and $r = 0.0033$ for $N_* = 60$; and $n_s = 0.959$ and $r = 0.0048$ for $N_* = 50$. We also calculated $n_s$ and $r$ numerically for each of the trajectories described above, and found $n_s = 0.967$ and $r = 0.0031$ for $N_* = 60$, and $n_s = 0.960$ and $r = 0.0044$ for $N_* = 50$. These values sit right in the most-favored region of the latest observations. (See Fig. 1 in \cite{PlanckInflation}.) Even for a low reheat temperature, we find $n_s$ within $2\sigma$ of the {\it Planck} value for $N_* \geq 38$. The predicted value $r \sim 10^{-3}$ could be tested by upcoming CMB polarization experiments. 

For the running of the spectral index, $\alpha \equiv d n_s / d \ln k$, we use Eq. (\ref{nsr}), the general relationship $\left( d x / d \ln k \right) \vert_* \simeq  \left( \dot{x} / H \right) \vert_*$ \cite{BTW}, and $N_* = N_{\rm tot} - \int_{t_i}^{t_*} H dt$ to find
\beq
\alpha = \frac{d n_s }{ d \ln k} \simeq  - \frac{2}{N_*^2} \left( 1 + \frac{3}{N_*} \right) ,
\label{alpha}
\eeq
which yields $\alpha = - 5.83 \times 10^{-4}$ for $N_* = 60$ and $\alpha = - 8.48 \times 10^{-4}$ for $N_* = 50$, fully consistent with the result from {\it Planck}, $\alpha = -0.0134 \pm 0.0090$, indicating no observable running of the spectral index \cite{PlanckInflation}. 

Meanwhile, for every trajectory in our large sample we numerically computed $f_{\rm NL}$ following the methods of \cite{KMS}. Across the whole range of couplings and initial conditions considered here, we found $\vert f_{\rm NL} \vert < 0.1$, consistent with the latest observations \cite{PlanckNonGauss}. In these models $f_{\rm NL}$ is exponentially sensitive to the fields' initial conditions, requiring a fine-tuning of ${\cal O} (10^{-4})$ to produce $\vert f_{\rm NL} \vert > 1$ \cite{KMS}. In the absence of such fine-tuning these models generically predict $\vert f_{\rm NL} \vert \ll {\cal O} (1)$.

Unlike several models with concave potentials analyzed in \cite{PlanckInflation}, multifield models with nonminimal couplings should produce entropy efficiently at the end of inflation, when $\xi_I (\phi^I )^2 < M_{\rm pl}^2$. The energy density and pressure are given by $\rho = {1 \over 2} \dot{\sigma}^2 + V (\phi^I)$ and $p = {1 \over 2} \dot{\sigma}^2 - V (\phi^I)$ \cite{KMS}. We confirmed numerically that for every trajectory in our large sample, the effective equation of state $w = p / \rho$ averaged to $0$ beginning at $t_{\rm end}$ (when $\epsilon = 1$) and asymptoted to $1/3$ within a few oscillations. This behavior may be understood analytically from the virial theorem, which acquires corrections proportional to gradients of the field-space metric coefficients, just like applications in curved spacetime \cite{VirialGR}. We find $\langle \dot{\sigma}^2 \rangle = \langle V_{, J} \varphi^J \rangle + \langle {\cal C} \rangle = \langle 2 M_{\rm pl}^4 V  / f \rangle + \langle {\cal C} \rangle$, where ${\cal C} \equiv - {1 \over 2} (\partial_J {\cal G}_{KL} ) \dot{\varphi}^K \dot{\varphi}^L \varphi^J$. More generally, inflation in these models ends with one or both fields oscillating quasi-periodically around the minimum of the potential, and hence preheating should be efficient \cite{BTW,Bassettmetric,Higgspreheating}.


The models in this class predict three basic possibilities for isocurvature perturbations, depending on whether inflation occurs while the fields are in a valley, on top of a ridge, or in a symmetric potential with $\lambda_I = g = \lambda$ and $\xi_I = \xi$ and hence no ridges (like Higgs inflation). The fraction $\beta_{\rm iso} (k) \equiv {\cal P}_{\cal S} (k) / [ {\cal P}_{\cal R} (k) + {\cal P}_{\cal S} (k) ] = T_{\cal SS}^2 / [1 + T_{\cal RS}^2 + T_{\cal SS}^2 ]$  \cite{PlanckInflation} may distinguish between the various situations. In each of these scenarios, $\omega^I \sim 0$ and hence $T_{\cal RS} \sim 0$. Inflating in a valley, $\eta_{ss} > 1$ so $\mu_s^2 / H^2 > 9/4$ and the (heavy) isocurvature modes are suppressed, $T_{\cal SS} \rightarrow 0$ and hence $\beta_{\rm iso} \sim 0$ for scales $k$ corresponding to $N_* = 60 - 50$. Inflating on top of a ridge, $\eta_{ss} < 0$ so $\mu_s^2 / H^2 < 0$ and the isocurvature modes grow via tachyonic instability, $T_{\cal SS} \gg 1$, and hence $\beta_{\rm iso} \sim 1$ across the same scales $k$. Scenarios in which the fields begin on top of a ridge and roll off at intermediate times can give any value $0 \leq \beta_{\rm iso} \leq 1$ depending sensitively upon initial conditions \cite{SSK}. In the case of symmetric couplings, $\mu_s^2 / H^2 \simeq 0$ \cite{GKS}, yielding $T_{\cal SS} \sim {\cal O} (10^{-3})$ and $\beta_{\rm iso} = 2.23 \times 10^{-5}$ for $N_* = 60$ and $\beta_{\rm iso} = 3.20 \times 10^{-5}$ for $N_* = 50$ \cite{Higgsisocurvfn}. 


Multifield models of inflation with nonminimal couplings possess a strong single-field attractor solution, such that they share common predictions for $n_s$, $r$, $\alpha$, $f_{\rm NL}$, and for efficient entropy production across a broad range of couplings and initial conditions. The predicted spectral observables provide excellent agreement with the latest observations. These models differ, however, in their predicted isocurvature perturbation spectra, which might help break the observational degeneracy among members of this class.

{\it Note}. While this paper was under review, similar results regarding attractor behavior in models with nonminimal couplings were presented in \cite{LindeAttractors}.

\acknowledgements{ It is a pleasure to thank Bruce Bassett, Rhys Borchert, Xingang Chen, Joanne Cohn, Alan Guth, Carter Huffman, Edward Mazenc, and Katelin Schutz for helpful discussions. This work was supported in part by the U.S. Department of Energy (DoE) under contract No. DE-FG02-05ER41360.}


\begin{thebibliography}{999}
\bibitem{unitarityfootnote} There has been vigorous debate over whether such models with $\xi_I \gg 1$ violate unitarity (see Refs. 12, 14, 15, 24, and 25 of \cite{GKS}). However, the constraint $H (t_*) / M_{\rm pl} < 10^{-5}$ means that the unitarity cutoff scale $\Lambda \gg H$ during inflation, and hence for the domain of interest we may rely on the classical background dynamics as analyzed here.
\bibitem{Higgsisocurvfn} In the case of Higgs inflation, in which the inflaton has no couplings other than to Standard Model particles, some as-yet unknown mechanism would be required to preserve primordial isocurvature perturbations through reheating. Primordial isocurvature perturbations could more readily survive reheating in symmetric-coupling models if the inflaton coupled to some non-Standard Model particles, such as cold dark matter particles.
\end{thebibliography}
\end{document}